\documentclass[3p,times,twocolumn]{elsarticle}

\usepackage{ecrc}

\volume{00}

\firstpage{1}

\journalname{Nuclear Physics B Proceedings Supplement}

\runauth{}

\jid{nuphbp}

\jnltitlelogo{Nuclear Physics B Proceedings Supplement}

\usepackage{amssymb,amsmath,float}

\usepackage[figuresright]{rotating}

\begin{document}

\begin{frontmatter}



\dochead{}

\title{A possible origin of gamma rays from the Fermi Bubbles}


\author{Satyendra Thoudam}
 \ead{s.thoudam@astro.ru.nl}
\address{Department of Astrophysics, IMAPP, Radboud University Nijmegen\\P.O. Box 9010, 6500 GL Nijmegen, The Netherlands}

\begin{abstract}
One of the most exciting discoveries of recent years is a pair of gigantic gamma-ray emission regions, the so-called ÒFermi bubblesÓ, above and below the Galactic center. The bubbles, discovered by the Fermi space telescope, extend up to $\sim 50^\circ$ in Galactic latitude and are $\sim 40^\circ$ wide in Galactic longitude. The gamma-ray emission is also found to correlate with radio, microwave and X-rays emission. The origin of the bubbles and the associated non-thermal emissions are still not clearly understood. Possible explanations for the non-thermal emission include cosmic-ray injection from the Galactic center by high speed Galactic winds/jets, acceleration by multiple shocks or plasma turbulence present inside the bubbles, and acceleration by strong shock waves associated with the expansion of the bubbles. In this paper, I will discuss the possibility that the gamma-ray emission is produced by the injection of Galactic cosmic-rays mainly protons during their diffusive propagation through the Galaxy. The protons interact with the bubble plasma producing $\pi^\circ$-decay gamma rays, while at the same time, radio and microwave synchrotron emissions are produced by the secondary electrons/positrons resulting from the $\pi^{\pm}$ decays.
\end{abstract}

\begin{keyword}
cosmic rays -- diffusion -- Galaxy -- gamma rays


\end{keyword}

\end{frontmatter}


\section{Introduction}
The Fermi space telescope has recently made an exciting discovery of two large gamma-ray emission regions, the so-called ÒFermi BubblesÓ, above and below the Galactic center \citep{Su2010}. The bubbles, whose origin still remains mysterious, extend up to $\sim 50^\circ$ in Galactic latitude and are $\sim 40^\circ$ wide in Galactic  longitude. They are coincident with microwave emission measured by WMAP and Planck satellite experiments \citep{Dobler2008, Planck2013}, share edges with X-rays emission measured by ROSAT telescope \citep{Snowden1997}, and also associate with two giant radio lobes discovered by S-PASS survey \citep{Carretti2013}.

The bubbles are most likely created by large energy injection from Galactic center in the past \citep{Su2010}, and therefore, their study provide valuable informations on the past activities of the Galaxy, particularly about the Galactic center region. Their proximity also give us a unique opportunity to understand similar extended lobes present in other galaxies. Moreover, a good understanding of their gamma-ray emission is crucial for indirect dark matter searches in the inner Galaxy, and will also improve our understanding of the cosmic-ray  population in the Galactic halo.

The non-thermal radiation from the Fermi Bubbles may result either from the interactions of high-energy cosmic-ray electrons with radiation/magnetic fields, or from inelastic collision of cosmic-ray nuclei with thermal nuclei. Some models suggest that the cosmic rays might originate from activities at the Galactic center and transported into the bubbles by high-speed winds or jets \citep{Crocker2011, Guo2012, Yang2012}, while others suggest they might be accelerated by shock waves or plasma turbulence present inside the bubbles \citep{Cheng2011, Mertsch2011, Fujita2013}.

It is not necessary that the process responsible for the formation of the bubbles also generates the cosmic rays that produce the non-thermal emission. The bubbles might have been created by activities at the Galactic center, and the cosmic rays might be injected into the bubbles by a different process. In this paper, I will discuss the possibility that the cosmic rays might be a population of Galactic cosmic rays which got injected into the bubbles during their propagation through the Galaxy. Considering that cosmic rays fill the entire volume of the Galaxy, I believe that this possibility can never be avoided. If successful, this model will provide a natural explanation for the origin of non-thermal radiations from the Fermi Bubbles.

\section{Some key features of the Fermi Bubbles}
The gamma-ray emission shows a flat intensity profile with sharp edges, and follows a hard energy spectrum of index $\sim -2$ in the range of $\sim 1-100$ GeV \citep{Su2010}. The flat intensity suggests a non-uniform cosmic-ray distribution that peaks towards the edge. The hard spectrum implies a cosmic-ray spectrum significantly harder than the equilibrium proton and electron spectra in the Galaxy. More recent analysis shows that the gamma-ray spectra above and below $\sim 10^\circ$ latitude are significantly different \citep{Hooper2013}. 

The microwave emission between $23-61$ GHz decreases sharply with increasing latitude below $\sim 35^\circ$, while above, the profile becomes similar to the gamma-ray profile and extends up to $50^\circ$ \citep{Planck2013}. The emission indicates the presence of a hard electron spectrum with index $\sim -2.1$. It is difficult to realize how high-energy electrons suffering severe radiative losses can maintain such a hard spectrum throughout the bubbles which extend up to $\sim 10$ kpc. 

The polarized radio emission between $2.3-23$ GHz is more extended up to $\sim 60^\circ$ latitude, and indicates a softer electron spectrum with index $\sim -3$ to $\sim -3.4$ steepening with increasing latitude \citep{Carretti2013}. This might indicate an additional electron population different from that producing the microwave emission.

The ROSAT X-rays map at $1.5$ keV shows limb brightening \cite{Snowden1997}, which agrees also with the recent SUZAKU measurements at $0.3$ keV \citep{Kataoka2013}. The X-rays emission is most likely thermal bremsstrahlung, although a synchrotron component from very high-energy electrons cannot be neglected. These complex multi-wavelength characteristics of the Fermi Bubbles present a tough challenge for any theoretical model. In this paper, I will primarily concentrate on the gamma ray emission.

\section{The model}
The model discussed here has been presented in detail in Thoudam (2013) \citep{Thoudam2013a}. After liberating from the sources, cosmic rays undergo diffusive propagation through the Galaxy. In stationary state in which the rate of production of cosmic rays is balanced by the loss mainly due to the leakage from the galaxy, it is assumed that cosmic rays fill the entire volume of the Galaxy. If the Fermi Bubbles do not contain any sources, the diffusive streaming of cosmic rays towards density gradient can result into a net flux of cosmic rays injected into the bubbles. Moreover, if the bubbles are expanding, there will be an additional CR injection as the bubbles sweep through the interstellar medium. The model assumes that the halo size of the Galaxy is large enough to contain the bubbles. Such a large halo is also preferred in order to explain the diffuse gamma-ray emissivity distribution in the Galaxy \citep{Ackermann2011}. 

The diffusive cosmic-ray injection flux into the bubbles follows $F_\mathrm{dif}=D_\mathrm{g}\nabla N_\mathrm{g}\propto D_\mathrm{g}N_\mathrm{g}$, where $\nabla N_\mathrm{g}$ represents the spatial density gradient of cosmic rays, and $N_\mathrm{g}$ and $D_\mathrm{g}$ represents the cosmic-ray density and cosmic-ray diffusion coefficient in the Galaxy. For protons, $N_\mathrm{g}$ is related to the cosmic-ray source spectrum $Q$ as $N_\mathrm{g}\propto Q/D_\mathrm{g}$ \citep{Thoudam2008}, therefore $F_\mathrm{dif}\propto Q$. The injection flux due to expansion of the bubbles is, $F_\mathrm{exp}=UN_\mathrm{g}\propto N_\mathrm{g}$, steeper than $F_\mathrm{dif}$, where $U$ is the expansion velocity. Thus, if $F_\mathrm{dif}>F_\mathrm{exp}$, the proton injection flux will follow the source spectrum. Once injected, cosmic rays undergo much slower diffusion than in the Galaxy due to high plasma turbulence inside the bubbles \citep{Yao2007}. At the same time, cosmic rays are also convected radially outward by the expanding plasma. If convection dominates, cosmic rays will have a distribution that peaks toward the edge of the bubble, in agreement with measurements. If the diffusion inside the bubble is energy-independent, as expected for turbulence generate by unstable winds \citep{Bykov1987}, the cosmic-ray energy spectrum will closely follow the injection spectrum.

Cosmic rays undergo inelastic collisions with the bubble plasma, and produce $\pi^\circ$-decay gamma rays and secondary electrons/positrons with spectra similar to the source cosmic rays in the Galaxy. The secondary electrons/positrons can produce synchrotron radiation in radio and microwave frequencies.

\section{Cosmic-rays inside the bubbles}
\subsection{Cosmic-ray distribution}
The evolution and distribution of cosmic ray density inside the bubbles can be described by the following time dependent diffusion-loss equation \citep{Thoudam2013a},
\begin{equation}
\frac{\partial}{\partial x}\left(D_\mathrm{b}\frac{\partial N_\mathrm{b}}{\partial x}\right)-\frac{N_\mathrm{b}}{\tau}=\frac{\partial N_\mathrm{b}}{\partial t},
\end{equation}
where $N_\mathrm{b} (x,E,t)$ represents the number density of cosmic rays with kinetic energy $E$ at a given time $t$ and position $x$. The position is measured from the edge of the bubble where $x>0$ represents the region inside the bubble, and $x<0$ represents the region outside. $D_\mathrm{b}$ is the cosmic-ray diffusion coefficient inside the bubbles, and $\tau=1/(n_\mathrm{b}v\sigma)$ is the inelastic collision time of cosmic rays with the bubble plasma where $n_\mathrm{b}$ is the plasma density, $v$ is  the cosmic-ray velocity, and $\sigma$ is the inelastic collision cross-section.

There are some evidence that unstable large-scale Galactic winds are present in the inner region of the Galaxy (e.g., \citep{Everett2008}). Such unstable winds can generate a turbulence wave spectrum inside the bubbles that follows $k^{-2}$ in the short wavelength regime $k\gg1/L$, where $k$ denotes the wave number and $L$ is the characteristic length of turbulence injection (e.g., \citep{Bykov1987}). If the turbulence is injected at scales of several parsecs which is much larger than the gyro-radii of cosmic rays relevant for producing the Fermi bubble gamma rays, the cosmic-ray diffusion coefficient is expected to be independent of energy. It is taken as $D_\mathrm{b}=K\times 10^{28}$ cm$^2$ s$^{-1}$, where $K$ is a constant, and is kept as a parameter in the study which will be determined based on the observed gamma-ray emission profile. Since this study is mainly concerned with cosmic rays of kinetic energies above $\sim 1$ GeV, the weak logarithmic energy dependence of $\sigma$ at high energies \citep{Kelner2006} will be neglected, and a constant cross-section of $\sigma=32$ mb will be considered in the study.

For cosmic rays continuously injected into the bubbles with flux $F(E)$, the distribution at time $t$ is obtained by solving Equation (1), and folding in the effect of adiabatic energy loss due to the spherical expansion of the bubbles \citep{Thoudam2013a}. The final solution is given below:
\begin{align}
N_\mathrm{b}(x,E,t)=&\frac{1}{\sqrt{\pi D_\mathrm{b}}}\int_0^t dt^\prime\;\frac{F(E^\prime)}{\sqrt{(t-t^\prime)}}\left(\frac{t}{t^\prime}\right)^{2/3}\nonumber\\
&\times\exp\left[\frac{-x^2}{4D_\mathrm{b}(t-t^\prime)}-\frac{(t-t^\prime)}{\tau}\right].
\end{align}

\subsection{Cosmic-ray injection}
The cosmic-ray injection flux $F$ can be written as
\begin{equation}
F=\left[D_\mathrm{g}\frac{dN_\mathrm{g}}{dx}+ UN_\mathrm{g}\right]_{x=0},\nonumber
\end{equation}
where the first term on the right hand side represents the injection due to the diffusive motion of cosmic rays in the Galaxy $F_\mathrm{dif}$, and the second term is due to the expansion of the bubbles in the interstellar medium $F_\mathrm{exp}$. They are calculated at the surface of the bubbles, i.e., at $x=0$. To estimate $F_\mathrm{dif}$, the cosmic-ray density gradient along the $z$ direction perpendicular to the Galactic plane will be first calculated. This is done as follows. The cosmic-ray density as a function of $z$ is given by \citep{Thoudam2008}
\begin{equation}
N_\mathrm{g}(z,E)\propto\frac{Q(E)}{D_\mathrm{g}(E)} f(z),
\end{equation}
where $Q(E)$ and $D_\mathrm{g}$ respectively represent the cosmic-ray source spectrum and the diffusion coefficient in the Galaxy, and $f(z)$ is a function that depends weakly on the cosmic-ray energy. The cosmic-ray density gradient then follows
\begin{equation}
\frac{dN_\mathrm{g}}{dz}\propto\frac{Q(E)}{D_\mathrm{g}}\frac{df}{dz}.
\end{equation}
From Equation (4), it can be seen that
\begin{equation}
D_\mathrm{g}\frac{dN_\mathrm{g}}{dz}\propto Q(E).
\end{equation}
Equation (5) shows that the diffusive injection flux $F_\mathrm{dif}$ will follow the source spectrum of cosmic rays in the Galaxy, as discussed in Section 3. The halo size of the Galaxy is taken to be $10$ kpc, large enough to contain the bubbles. For this value of halo size, the value of $D_\mathrm{g}$ is obtained as  $D_\mathrm{g}=D_0 (E/3\mathrm{GeV})^{0.6}$ with $D_0=6\times 10^{28}$ cm$^2$ s$^{-1}$, based on the measured boron-to-carbon ratio. 

 $F_\mathrm{exp}$ can be calculated if the expansion velocity of the bubbles is known. The velocity is estimated using data from the combined \textit{Planck-WMAP} measurements of the microwave emission from the region. The \textit{Planck-WMAP} data do not show any break or steepening in the emission spectrum within the frequency range of $23-61$ GHz. This suggests that the synchrotron loss time of the electrons emitting the microwave emission is larger than the age of the bubbles. Assuming that the electrons radiate at critical frequencies, the highest measured frequency of $61$ GHz corresponds to an electron energy of $65.6$ GeV for a magnetic field value of $1.3 \mu$G  inside the bubbles. The magnetic field is calculated using the relation $B(z)=7e^{-z/3\mathrm{kpc}}$ $\mu \mathrm{G}$ used in the {\footnotesize{GALPROP}} cosmic-ray  propagation code \citep{Strong2010}, and by taking $z=5$ kpc which is taken to be the position of the bubble center based on the observed geometry of the bubbles. This consideration implies that the age of the bubbles must be less than $1.1\times 10^8$ yr. This gives a lower limit of the expansion velocity of the bubbles at $39.6$ km s$^{-1}$ for the present radius of the bubble which is taken to be $4.5$ kpc.

In reality, the injection flux may vary for different positions and also for different directions in the Galaxy. The study will neglect such possible variations, and assumes a uniform injection that corresponds to the injection flux at $z=5$ kpc. The total injection flux is then taken as $F=C\left[F_\mathrm{dif}+F_\mathrm{exp}\right]_{z=5\mathrm{kpc}}$, where $C$ is a constant, hereafter referred to as the injection fraction, which is introduced in order to take care of the unknown actual fraction of cosmic rays injected into the bubbles. It is  further assumed that the cosmic-ray diffusion coefficient scales inversely with the magnetic field strength in the Galaxy as $D_0(z)=6\times 10^{28} e^{z/3\mathrm{kpc}}$ cm$^2$ s$^{-1}$. This gives a value of $D_0=31.7\times 10^{28}$ cm$^2$ s$^{-1}$ at $z=5$ kpc. For this value of $D_0$ and $U=39.6$ km s$^{-1}$, the injection flux is found to be dominated by $F_\mathrm{dif}$ for cosmic-ray energies above $1$ GeV. For larger values of $U$, $F_\mathrm{exp}$ may become important particularly at lower energies, and this will be discussed later in Section 6.

The cosmic-ray source spectrum is chosen to be a broken power law with indices $\Gamma=2.2$ and $2.09$ at energies below and above $300$ GeV. This form of source spectrum is chosen so as to reproduce the recently  measured proton spectrum by various experiments such as the ATIC \citep{Panov2007}, CREAM \citep{Yoon2011}, and PAMELA, \citep{Adriani2011} which exhibit a spectral hardening above $\sim 250$ GeV. The observed hardening might be due to the effect of local sources, propagation effects or signature of the cosmic-ray source spectrum itself (see e.g., Refs. \citep{Thoudam2012, Thoudam2013b, Thoudam2014} and references therein).

\section{Gamma-ray emission from the bubbles}
In the present model, the Fermi Bubble $\gamma$-rays are the decay products of $\pi^0$ mesons produced from the inelastic collision of cosmic rays with the bubble plasma. For a uniform plasma density inside the bubbles, the $\gamma$-ray emissivity is expected to follow the distribution of cosmic rays. The $\gamma$-ray emissivity at a given time and as a function of the radial coordinate $r$ measured from the bubble center is calculated as
\begin{equation}
q_\gamma(r,E_\gamma,t)=2\int_{E^\mathrm{min}_{\pi^0}}^\infty \frac{q_{\pi^0}(r,E_{\pi^0},t)}{\sqrt{E^2_{\pi^0}-m^2_{\pi^0}}}dE_{\pi^0},
\end{equation}
where $E_\gamma$ denotes the $\gamma$-ray energy, $E_{\pi^0}$ and $m_{\pi^0}$ denote the total and the rest mass energy of the pion respectively, and $E_{\pi^0}^\mathrm{min}=E_\gamma+m_{\pi^0}^2/4E_\gamma$. The $\pi^0$ mesons emissivity is given by
\begin{equation}
q_{\pi^0}(r,E_{\pi^0},t)=\frac{\tilde{n}}{\tilde{k}}cn_\mathrm{b}\sigma N_\mathrm{b}(r,T,t)
\end{equation}
where $c$ is the velocity of light, and $N_\mathrm{b}(r,T,t)$ is the radial dependent cosmic-ray proton density obtained from Eq. (2) by replacing $x$ with $R-r$ and writing as a function of the total energy $T$ which is related to the pion total energy $E_{\pi^0}$ and the proton rest mass energy $m_\mathrm{p}$ as $T=m_\mathrm{p}+E_{\pi^0}/\tilde{k}$. The values of $\tilde{k}$ and $\tilde{n}$ are taken as $\tilde{k}=0.17$ and $\tilde{n}=1$ from Ref. \cite{Kelner2006}, and a constant inelastic cross-section of $\sigma=32$ mb will be assumed. The $\gamma$-ray intensity in a given direction is calculated as
\begin{equation}
I_\gamma(E_\gamma)=\frac{1}{4\pi}\int^{y_2}_{y_1} q_\gamma(y,E_\gamma,t)dy
\end{equation}
where the integration is along the line of sight distance $y$, and the integration limits are given by the intersection points of the line of sight with the bubble surface.

\section{Results}
The calculation will be performed by taking $t=1.1\times 10^8$ yr which is the upper limit of the bubble age that has been determined, $U=39.6$ km s$^{-1}$ which is the lower limit of the expansion velocity, and $n_\mathrm{b}=3\times 10^{-3}$ cm$^{-3}$. This value of $n_\mathrm{b}$ is the averaged value in the region taken from Ref. \cite{Everett2008} that explains the diffuse soft X-ray emission. Once the values of $t, U$ and $n_\mathrm{b}$ are fixed, the distribution of cosmic rays inside the bubbles is determined by the value of $D_\mathrm{b}$. This is determined by choosing a value of $K$ that produces the best agreement between the model gamma-ray intensity profile and the measured data. The best fit value is found to be $K=0.26$, and the resulting intensity profiles are shown in Figure 1 along with the measured data for two different energy regions $1-5$ GeV and $5-20$ GeV.
\begin{figure}[h]
\hspace{-7mm}
\includegraphics*[width=0.53\textwidth,angle=0,clip]{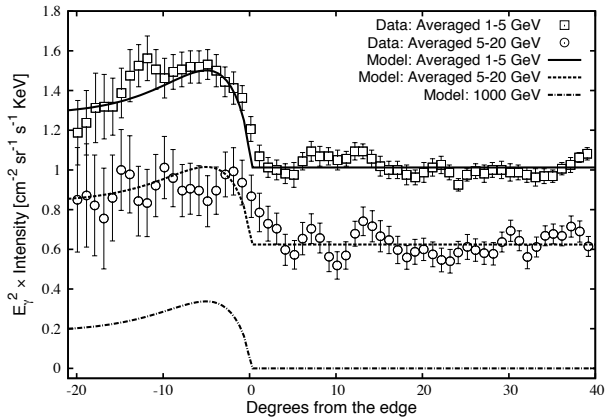}
\caption{\label {fig1} Model prediction of the projected $\gamma$-ray intensity profile for the averaged $1-5$ GeV (solid line) and $5-20$ GeV (dashed line) energy regions. Also shown is the predictions for the $1000$ GeV energy (dot-dashed line). The results are calculated by taking $K=0.26$, $t=1.1\times 10^8$ yr, $U=39.6$ km s$^{-1}$, and $n_\mathrm{b}=3\times 10^{-3}$ cm$^{-3}$. Data corresponds to the Southern Fermi bubble taken from Ref. \cite{Su2010}.}
\end{figure}
In Figure 1, $0^\circ$ represents the edge of the bubbles, and the negative and positive angles represent the regions inside and outside the bubbles respectively. The model predictions are added with backgrounds obtained by fitting the distribution between $5^\circ$ and $40^\circ$. It can be seen that the model reproduces the measured data quite well. Also shown in Figure 1 is the prediction for the $1000$ GeV energy which can be tested in future. The result at high energy is very similar to those at low energies because of the energy independent nature of the diffusion coefficient inside the bubbles. It can be noted that the result at high energy shown in Figure 1 is clearly different from that expected in the leptonic stochastic acceleration model presented in Ref. \cite{Mertsch2011} which predicted a significant limb brightening at high energies.

\begin{figure}[h]
\hspace{-7mm}
\includegraphics*[width=0.53\textwidth,angle=0,clip]{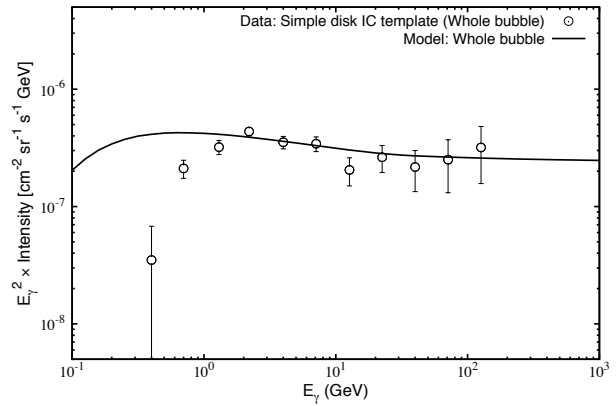}
\caption{\label {fig1} Gamma-ray spectra for a whole bubble (solid line). The calculation assumes $C=0.8$, and a cosmic-ray source index of $\Gamma=2.2$ and 2.09 below and above $300$ GeV in the Galaxy. Data are from Ref. \cite{Su2010}.}
\end{figure}

Figure 2 shows a comparison between the calculated gamma-ray spectrum over a whole bubble and the measured data taken from Ref. \citep{Su2010}. The model spectrum is normalized to the data at $4$ GeV, and that requires a cosmic-ray injection fraction of $C=0.8$. Note that the same value of cosmic-ray injection fraction has been used in Figure 1. In Figure 2, the model prediction seems to agree nicely with the data above $\sim 1$ GeV where the measurement uncertainties are small. However, below $\sim 1$ GeV, the data show a sharp turn over which cannot be explained satisfactorily by the present model.

A pure diffusion model of cosmic-ray propagation has been adopted in the calculation of the cosmic-ray spectrum in the Galaxy. It is interesting to see that the same source index, required to reproduce the measured proton spectrum in the pure diffusion model, also reproduces the gamma-ray spectrum from the Fermi Bubbles. This indicates that the high energy particles responsible for the gamma-ray production from the Fermi Bubbles might indeed be some fraction of Galactic cosmic rays which got injected into the bubbles during their diffusive propagation through the Galaxy. However, it can be noted that propagation models which are based on reacceleration of cosmic rays in the Galaxy, that requires a steeper cosmic-ray source index of $\Gamma\sim 2.4$, will be difficult to explain the gamma-ray spectrum from the bubbles.

\begin{figure}
\hspace{-7mm}
\includegraphics*[width=0.53\textwidth,angle=0,clip]{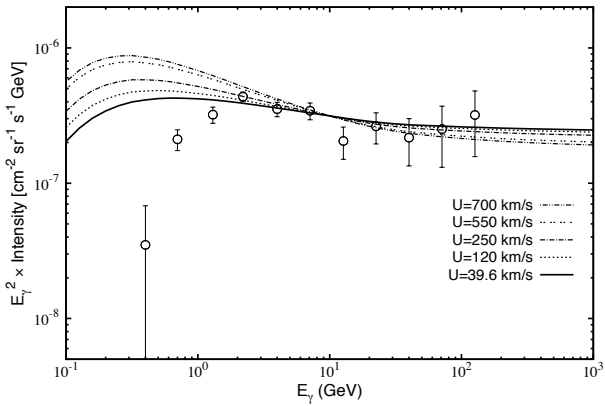}
\caption{\label {fig1} Gamma-ray spectra for a whole bubble for different values of expansion velocity in the range of $U=39.6-700$ km s$^{-1}$. Data are taken from Ref. \cite{Su2010}.}
\end{figure}

The results presented in Figures $1$ and $2$ assume the lower limit of expansion velocity deduced from the  {\it{Planck}}-WMAP data. Choosing a higher expansion velocity will increase the contribution of $F_\mathrm{exp}$ resulting into a steeper cosmic-ray injection spectrum, and ultimately into a steeper $\gamma$-ray spectrum. This is depicted in Figure 3, where gamma-ray spectra for a whole bubble corresponding to different expansion velocities in the range of $U=39.6-700$ km s$^{-1}$ are shown. The spectra are normalized to the case of $U=39.6$ km s$^{-1}$ at $10$ GeV. In a detailed study given in Thoudam 2013 \citep{Thoudam2013a}, it has been shown explicitly that the model prediction becomes inconsistent with the measured data above $1$ GeV at $U\geq 180$ km s$^{-1}$. This sets an upper limit on the expansion velocity of the bubbles at $U<180$ km s$^{-1}$, which corresponds to a lower limit of the bubble age at $t_{\mathrm{age}}>2.44\times 10^7$ yr. This implies a synchrotron break in the underlying electron spectrum at energy below $298$ GeV, and also a corresponding break in the synchrotron emission spectrum at frequency below $1258$ GHz. This can be checked by sensitive measurements in future.

\section{Conclusions}
A possible explanation for the gamma-ray emission from the Fermi Bubbles has been presented. Unlike other existing models, the model presented here does not invoke any additional sources or particle production sources other than those responsible for the production of the bulk of the galactic cosmic rays. It has been shown that the gamma-rays from the bubbles are produced by a population of Galactic cosmic-ray protons which got injected into the bubbles during their diffusive propagation through the Galaxy. The injected cosmic rays interact with  matter inside the bubbles producing $\pi^\circ$-decay gamma rays. The secondary electron/positrons resulting from the same interaction process produce synchrotron emission which might explain the observed microwave and radio emissions from the bubbles.
\\
\\
{\textit{Acknowledgements:}} I wish to thank the organizers for such a wonderful conference at such a beautiful location.









\end{document}